\pdfoutput=1 
\documentclass{nature}

\usepackage{multirow}
\usepackage{ccaption}
\usepackage{subcaption}
\usepackage{graphicx}
\usepackage{lipsum}
\usepackage{ragged2e}
\usepackage{fixltx2e}
\usepackage{hyperref}% to split url multiple lines

\RequirePackage{mathpazo}
\usepackage{amsmath}

\usepackage{comment}
\usepackage{cite}
\usepackage{adjustbox} % to adjust the size of table

\makeatletter
\newenvironment{figurehere}
{\def\@captype{figure}}
{}
\makeatletter

\usepackage{xcolor}

\title{\begin{center}  Paving the Way for Pass Disturb Free Vertical NAND Storage via A
Dedicated and String-Compatible \\ Pass Gate \end{center}}

\author
{ \centering
{%\setstretch{1.5}
\large{Zijian Zhao$^{1}$, Sola Woo$^{2}$, Khandker Akif Aabrar$^{2}$, Sharadindu Gopal Kirtania$^{2}$, Zhouhang Jiang$^{1}$, Shan Deng$^{1}$, Yi Xiao$^{3}$, Halid Mulaosmanovic$^{4}$, Stefan Duenkel$^{4}$, Dominik Kleimaier$^{4}$, Steven Soss$^{4}$, Sven Beyer$^{4}$, Rajiv Joshi$^{5}$, Scott Meninger$^{6}$, Mohamed Mohamed$^{6}$, Kijoon Kim$^{7}$, Jongho Woo$^{7}$, Suhwan Lim$^{7}$, Kwangsoo Kim$^{7}$, Wanki Kim$^{7}$, Daewon Ha$^{7}$, Vijaykrishnan Narayanan$^{3}$, Suman Datta$^{2}$, Shimeng Yu$^{2}$, and Kai Ni$^{1\dagger}$\\}
\vspace{3ex}
\normalsize{$^{1}$University of Notre Dame, Notre Dame, IN 46556, USA}\\
\normalsize{$^{2}$Georgia Institute of Technology, Atlanta, GA 30332, USA}\\
\normalsize{$^{3}$Pennsylvania State University, State College, PA 16802, USA}\\
\normalsize{$^{4}$GlobalFoundries Fab1 LLC \& Co. KG, 01109 Dresden, Germany}\\
\normalsize{$^{5}$IBM Thomas J. Watson Research Center, Yorktown Heights, NY 10598, USA}\\
\normalsize{$^{6}$MIT Lincoln Laboratory, Lexington, MA 02421, USA}\\
\normalsize{$^{7}$Samsung Electronics Co., Ltd, Hwaseong, Gyeonggi 18448, South Korea}\\
\vspace{2ex}
\normalsize{$^{\dagger}$To whom correspondence should be addressed} \\
\normalsize{Email: kni@nd.edu} \\
}}

\begin{document}
\flushbottom
\maketitle
\vspace{3ex}
\begin{abstract}
In this work, we propose a dual-port cell design to address the pass disturb in vertical NAND storage, which can pass signals through a dedicated and string-compatible pass gate. We demonstrate that: i) the pass disturb-free feature originates from weakening of the depolarization field by the pass bias at the high-\textit{V}\textsubscript{TH} (HVT) state and the screening of the applied field by channel at the low-\textit{V}\textsubscript{TH} (LVT) state; ii) combined simulations and experimental demonstrations of dual-port design verify the disturb-free operation in a NAND string, overcoming a key challenge in single-port designs; iii) the proposed design can be incorporated in a highly scaled vertical NAND FeFET string and the pass gate can be incorporated into the existing 3D NAND with the negligible overhead of the pass gate interconnection through a global bottom pass gate contact in the substrate.
\end{abstract}

\thispagestyle{empty}

%\noindent Please note: Abbreviations should be introduced at the first mention in the main text – no abbreviations lists. Suggested structure of main text (not enforced) is provided below.

\section*{\large Introduction}
Vertical NAND flash storage, as shown in Fig.\ref{fig:Fig1}\textbf{a}, has been the backbone of current digital storage system, due to its high density and low cost \cite{compagnoni2017reviewing, jang2009vertical, goda2021recent, kim2009novel}. It attains an impressive memory density by vertically stacking memory layers, as shown in the cross-sectional schematic in Fig.\ref{fig:Fig1}\textbf{b}, with the state-of-the-art design surpassing 200-300 layers\cite{meyer20223d,skhynixwebsite} and marching towards an unprecedented 1000 layers \cite{goda20203}. Concurrently, there is a growing interest in exploring cell technologies beyond conventional flash transistors to lower operation voltage and enhance operational speed. One noteworthy contender is the ferroelectric field-effect transistor (FeFET), gaining attention following the discovery of ferroelectricity in thin doped HfO\textsubscript{2} films \cite{boscke2011ferroelectricity,schroeder2022fundamentals, florent2017first}. Possessing a similar transistor structure, vertical FeFET demonstrates the potential for multi-bit storage through partial polarization switching and an exceptionally efficient polarization switching process \cite{zeng2021electric, zeng20192, chatterjee2022comprehensive}, making it a promising candidate for mass storage. However, whether employing commercialized flash or potential FeFET storage, a common challenge faced by the vertical NAND structure is the escalating disturbance to memory states during array operation under aggressive scaling \cite{mulaosmanovic2020impact}. Such disturbance could increase the bit error rate and under extreme cases, cause information loss \cite{micheloni2017array}. In this study, we thoroughly investigate disturbance issues in NAND storage, using the FeFET as an illustrative platform. To address these challenges, we introduce a novel dual-port FeFET structure, as shown in Fig.\ref{fig:Fig1}\textbf{c}, designed to enable disturbance-free operation.

Due to its unique serial structure, vertical NAND array is more susceptible to disturb than other structures of memory array during memory read and write operations \cite{prince2014vertical}. For example, to read the target cell in a conventional single-port vertical NAND string, where the program and pass biases are applied on the same gate (i.e., word line, (WL)), other unselected cells in the same string need a large pass voltage, \textit{V}\textsubscript{PASS}, on the gate to pass string end voltages to the target cell \cite{aritome2015nand}. For this purpose, \textit{V}\textsubscript{PASS} needs to be greater than the highest threshold voltage (\textit{V}\textsubscript{TH}) to ensure turning ON unselected cells regardless of their \textit{V}\textsubscript{TH} states. As a result, a high \textit{V}\textsubscript{PASS} could disturb the memory states. In addition, the choice of \textit{V}\textsubscript{PASS} is even more delicate during memory write operation in order to balance between various disturbs \cite{micheloni2010inside}. Fig.\ref{fig:Fig1}\textbf{d} shows a typical global self-boosting program inhibition scheme adopted in vertical NAND \cite{suh19953,kang2017natural}. After a block erase, the target cells on the selected page will be programmed while other cells on the same page will be inhibited. This is achieved by applying a ground and \textit{V}\textsubscript{CC} on the selected cells and unselected cells, respectively. In this way, the selected cells have enough voltage applied to program the state while unselected strings are floating and the channel potential will be raised by the applied voltages to inhibit programming. Fig.\ref{fig:Fig1}\textbf{e} shows the equivalent circuit model for the selected string, where the green resistors represented ON cells. Under this scenario, a high/low \textit{V}\textsubscript{PASS} will introduce severe pass/program disturb, respectively, as shown in Fig.\ref{fig:Fig1}\textbf{f}. When \textit{V}\textsubscript{PASS} is very high, the pass disturb to the cells on the same selected strings, as shown in Fig.\ref{fig:Fig1}\textbf{d}, will be significantly disturbed. However when \textit{V}\textsubscript{PASS} is very low, the unselected string channel potential may not be elevated enough to inhibit program disturb to unselected cells on the same page, as shown in Fig.\ref{fig:Fig1}\textbf{d}. 
Therefore, only a narrow \textit{V}\textsubscript{PASS} margin is available and sometimes a tradeoff has to be made. 

\begin{figurehere}
  \centering
    \includegraphics[scale=1,width=0.9\textwidth]{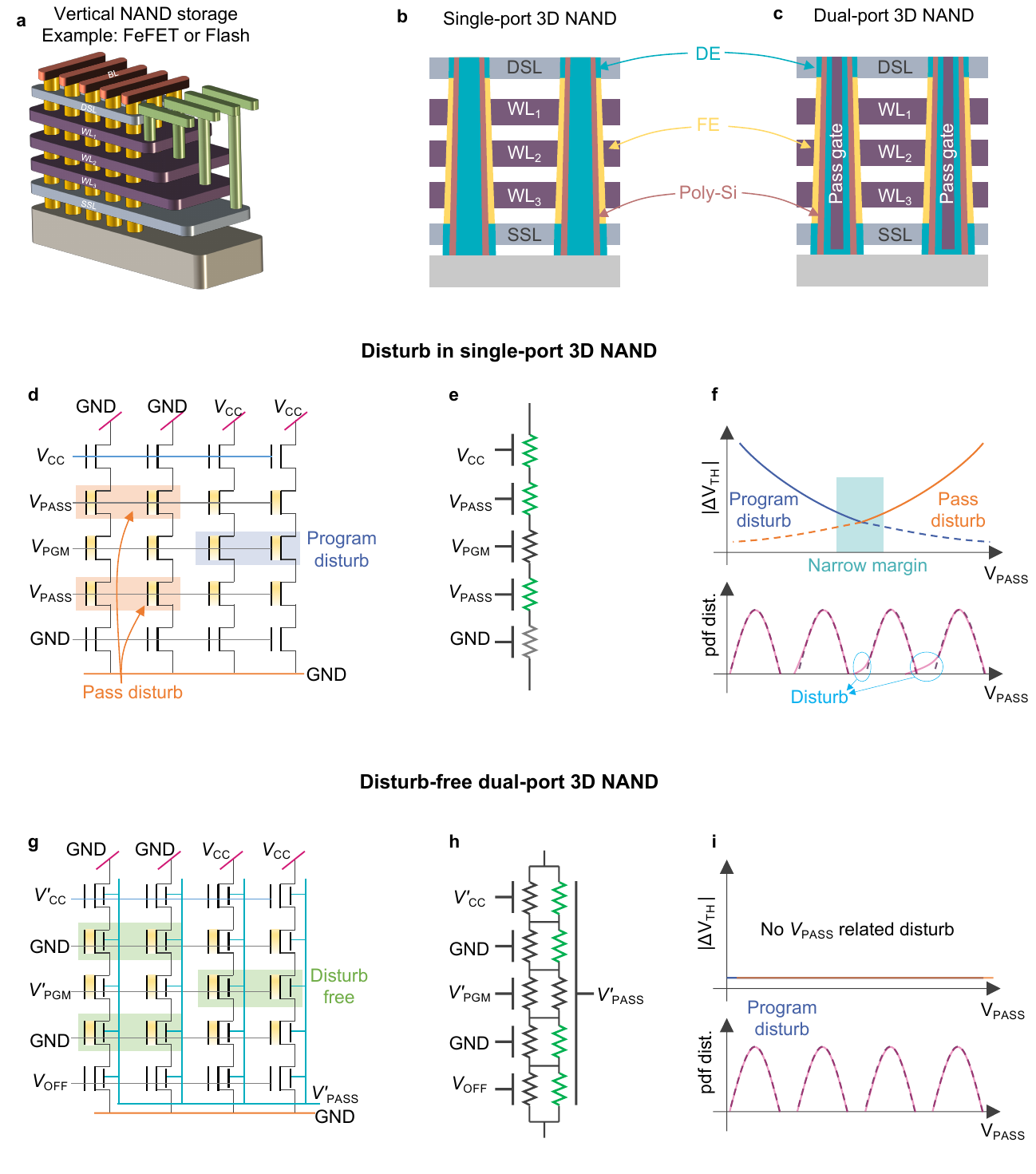}
    \captionsetup{parbox=none} % for caption to be split
    \caption{\textit{\textbf{Dual-port vertical NAND FeFET for pass disturb-free operation.} \textbf{a.} The conventional vertical NAND flash storage has been widely used as a high-density low-cost storage. FeFET can utilize the similar structure to achieve lower operation voltage and faster operation speed. \textbf{b.} The conventional vertical storage stacks hundreds of layers for a higher density, but suffers from the escalating disturbance due to the aggressive scaling. \textbf{c.} Adding a second port specifically for the read and pass operation can address these challenges. \textbf{d.} A typical global self-boosting program inhibition scheme for the program and inhibition operations. The program disturb occurs on the selected page while the pass disturb is seen on the unselected pages. \textbf{e.} The equivalent circuit model for the selected string. The green resistors represent ON cells. \textbf{f.} A high \textit{V}\textsubscript{PASS} causes the pass disturb while a low \textit{V}\textsubscript{PASS} increases the program disturb. The disturb-free margin is narrow. \textbf{g.} The proposed design features an independent non-ferroelectric pass gate for the read and pass operation. The \textit{V}\textsubscript{PASS} will no longer be applied to the ferroelectric gate, resulting in a pass disturb-free operation. \textbf{h.} The equivalent circuit model for the selected string. The pass transistors are turned ON from the pass gate. \textbf{i.} illustrates the negligible \textit{V}\textsubscript{TH} shifts and the related distribution.}}
    \label{fig:Fig1}
\end{figurehere}

To address this challenge, a dual-port vertical NAND, shown in Fig.\ref{fig:Fig1}\textbf{c}, is proposed in this work, which features an independent pass gate (PG) in the core of a NAND string and normal non-ferroelectric gate dielectric, which can turn ON the string channel for the passing operation. In this way, such a structure is compatible with the vertical NAND array, without incurring significant overhead when introducing the additional gate. Fig.\ref{fig:Fig1}\textbf{g} shows the programming bias scheme for the proposed dual-port vertical NAND array, where the pass operation is conducted using a dedicated pass gate. In this case, unselected cells on the same target strings will no longer need a high \textit{V}\textsubscript{PASS} on the ferroelectric gate, thus completely eliminating the pass disturb. Fig.\ref{fig:Fig1}\textbf{h} shows the equivalent circuit model for this case, where the channel is turned ON from the back such that the string end voltages can be passed to the target cells. As a result, we argue that there will be negligible disturb to the memory states with the dual-port transistor design and the array distribution will be highly robust against the pass disturb, as illustrated in Fig.\ref{fig:Fig1}\textbf{i}. In the following, we will clarify the origin of disturb-free operation in the dual-port FeFETs and validate experimentally in both the front-end-of-line (FEOL) and back-end-of-line (BEOL) HfO\textsubscript{2} based FeFETs, demonstrate disturb free operation in NAND FeFET string, and perform technology computer-aided design (TCAD) studies on scaled vertical FeFET.

\section*{\large Dual-Port FeFET Enabling Disturb-Free Operation}
The origin of pass disturb in the single-port FeFET and the disturb-free operation in dual-port FeFET is illustrated in Fig.\ref{fig:Fig2}. For a single-port FeFET, where the write/read/pass biases are all applied on the gate with the ferroelectric film, disturb can be resulted. If the single-port FeFET is in the low-\textit{V}\textsubscript{TH} (LVT) state, as shown in Fig.\ref{fig:Fig2}\textbf{a}, a \textit{V}\textsubscript{PASS} applied on the gate generates an applied electric field, \textit{E}\textsubscript{APP}, which is aligned with the polarization, thus weakening the depolarization field (\textit{E}\textsubscript{DEP}) and improving the state stability. In contrast, when the FeFET is in the high-\textit{V}\textsubscript{TH} (HVT) state, as shown in Fig.\ref{fig:Fig2}\textbf{b}, the applied \textit{E}\textsubscript{APP} exerted by the \textit{V}\textsubscript{PASS} will be against the polarization, thus enhancing the \textit{E}\textsubscript{DEP} and causing retention loss. Pass disturb to other intermediate states will also be between that for the LVT and HVT extreme states. Such pass disturb can be eliminated through structural modification by adopting a dual-port FeFET, where a second electrical gate with non-ferroelectric dielectric is incorporated \cite{mulaosmanovic2021dualport}. Due to the thin film channel, the polarization set through the ferroelectric gate controls the channel carrier concentration, which can also be sensed through the non-ferroelectric gate. This innate structural properties ensures pass disturb free operation. For example, when the FeFET is set to the LVT state, as shown in Fig.\ref{fig:Fig2}\textbf{c}, the pass bias applied on the non-ferroelectric pass gate could cause an \textit{E}\textsubscript{APP} against the polarization. However, in this case, the channel is fully turned ON such that all the applied bias is screened by the channel electrons and almost no electric field can penetrate through the channel and reach the ferroelectric layer. As a result, the ferroelectric state is retained. For HVT state, the \textit{E}\textsubscript{APP} exerted by the pass bias is aligned with the polarization, thus helping HVT state retention, as shown in Fig.\ref{fig:Fig2}\textbf{d} \cite{zhao2023powering}.

To validate the disturb-free pass operation in dual-port FeFET, two types of FeFETs are integrated and characterized, one FEOL version and one BEOL version. The FEOL dual-port FeFET is simply realized with fully depleted silicon-on-insulator (FDSOI) FeFET, as shown in Fig.\ref{fig:Fig2}\textbf{e}, where the ferroelectric sits on the top of the Si channel and the buried oxide (BOX) is the non-ferroelectric dielectric while the p-well contact can serve as the pass gate. Fig.\ref{fig:Fig2}\textbf{f} shows the transmission electron microscopy (TEM) image of the testing device integrated on the 22 nm FDSOI platform \cite{dunkel2017fefet}. First, theoretical understanding of the disturb-free pass operation is gained by checking the ferroelectric electric field at different pass biases for both single-port and dual-port FEOL FeFET using a well-calibrated TCAD models of the FDSOI transistor. Fig.\ref{fig:Fig2}\textbf{g} and \textbf{h} shows the ferroelectric electric field for the HVT state and LVT state, respectively. It shows that the depolarization field got enhanced in the single-port device while reduced in the dual-port device at the HVT state, thus supporting the physical picture discussed above. For the LVT state, the depolarization field is reduced for the single-port device while remains constant for the dual-port device due to the channel screening. Experimental validation on the FDSOI FeFET is also conducted. Fig.\ref{fig:Fig2}\textbf{i} and \textbf{j} show measured \textit{I}\textsubscript{D}-\textit{V}\textsubscript{G} curves swept on the ferroelectric gate and the non-ferroelectric gate, respectively. For each gate sweep, it shows the classical behavior that the \textit{V}\textsubscript{TH} can be tuned linearly with the other gate bias. It also shows a much larger memory window for the non-ferroelectric gate sweep due to the much larger equivalent oxide thickness (EOT) \cite{mulaosmanovic2021dualport}. Fig.\ref{fig:Fig2}\textbf{k} shows the disturb to the HVT and LVT state on a single-port FeFET when \textit{V}\textsubscript{PASS} is applied on the write gate (WG). It shows a severe disturb to the HVT state by \textit{V}\textsubscript{PASS} in a single-port FeFET and no disturb to the LVT state. However, for the dual-port FeFET, as shown in Fig.\ref{fig:Fig2}\textbf{l}, both the HVT and LVT states are stable under \textit{V}\textsubscript{PASS} bias, therefore verifying the disturb-free pass operation. The pass-disturb free operation is also experimentally verified on the BEOL dual-port FeFET. The FeFET is realized with the ferroelectric placed below the amorphous metal oxide thin film channel (e.g., tungsten doped indium oxide (IWO) in this work) and the non-ferroelectric layer placed above the channel, as shown in the Supplementary Fig.\ref{fig:sups1}.

\begin{figurehere}
  \centering
    \includegraphics[scale=1,width=0.95\textwidth]{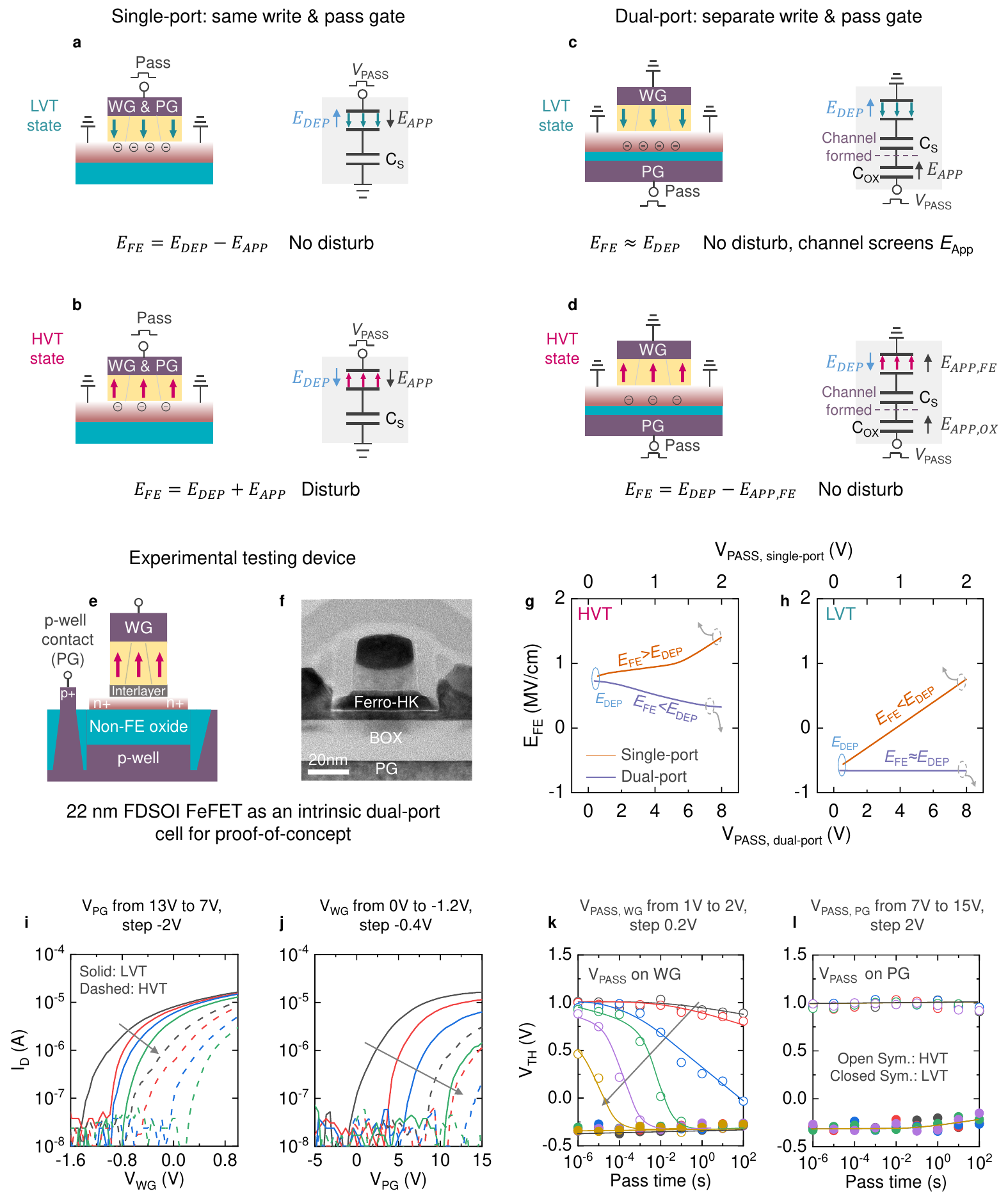}
    \captionsetup{parbox=none} % for caption to be split
    \caption{\textit{\textbf{Origin of pass disturb in the single-port FeFET and the pass disturb free in the dual-port FeFET.} \textbf{a.} For the single-port FeFET in the LVT state, the \textit{V}\textsubscript{PASS} applied to the gate with the ferroelectric film generates \textit{E}\textsubscript{APP}, which counteracts the \textit{E}\textsubscript{DEP}. \textbf{b.} In contrast, the \textit{E}\textsubscript{APP} applied in the HVT state enhances \textit{E}\textsubscript{DEP}, causing pass disturb. \textbf{c.} Dual-port FeFET offers two independent paths for write and read/pass operations, respectively. For the FeFET in LVT state, although the \textit{E}\textsubscript{APP} aligns with the \textit{E}\textsubscript{DEP}, the applied pass bias is screened by the formed channel electrons, resulting in no enhancement in \textit{E}\textsubscript{DEP}. \textbf{d.} For the FeFET in the HVT state, the \textit{E}\textsubscript{APP} is aligned with the polarization, thus enhancing retention. \textbf{e-f.} The schematic and the TEM of the FDSOI FeFET used for experimental verification \cite{dunkel2017fefet}. \textbf{g-h.} The simulated results of the \textit{E}\textsubscript{FE}-\textit{V}\textsubscript{PASS} curves confirm the analysis of the read disturb issue. \textbf{i-j.} \textit{I}\textsubscript{D}-\textit{V}\textsubscript{G} curves for WG read and PG read measured after $\pm$4 V, 1 $\mu$s write pulses. \textbf{k-l.} \textit{V}\textsubscript{TH} shifts measured after different pass time lengths under different pass voltages applied on the WG and PG, respectively. No \textit{V}\textsubscript{TH} shift is found when measuring with PG.}}
    \label{fig:Fig2}
\end{figurehere}

\section*{\large Pass Disturb Free Operation in NAND FeFET String}
Next, the pass disturb free operation in a NAND FeFET string will be tested. Without loss of generality, a string composed of three FDSOI FeFETs is tested, which contains all the key features of NAND array and is enough to demonstrate the working principles. First, pass disturb in single-port FeFET NAND string is demonstrated. Fig.\ref{fig:Fig3}\textbf{a} shows the experimental setup and waveforms for single-port FeFET based NAND string. For the ease of characterization, during the experiment, gates of the top transistor (i.e., T\textsubscript{1}) and bottom transistor (i.e., T\textsubscript{3}) are wired together, which makes i.e., T\textsubscript{1} and i.e., T\textsubscript{3} have the same state and operations at all times. The waveform shows a case where \textit{V}\textsubscript{READ}/\textit{V}\textsubscript{PASS} are applied to selected (T\textsubscript{1}\&T\textsubscript{3})/unselected (T\textsubscript{2}) transistors. With this setup, the \textit{I}\textsubscript{D}-\textit{V}\textsubscript{G} characteristics of T\textsubscript{1}\&T\textsubscript{3} can be obtained, as shown in Fig.\ref{fig:Fig3}\textbf{b}, demonstrating successful operation of the string. Similarly, when the selected cell is T\textsubscript{2}, a \textit{V}\textsubscript{PASS} will be applied on T\textsubscript{1}\&T\textsubscript{3} and \textit{I}\textsubscript{D}-\textit{V}\textsubscript{G} characteristics of T\textsubscript{2} can be measured, as shown in Fig.\ref{fig:Fig3}\textbf{c}. In this case, irrespective of the states of unselected cells, the correct memory information of selected cell can be successfully sensed. Following this demonstration, the pass disturb is characterized, as shown in Fig.\ref{fig:Fig3}\textbf{d}, where \textit{V}\textsubscript{PASS} is applied on T\textsubscript{2} for T\textsubscript{1}\&T\textsubscript{3} sensing. Different \textit{V}\textsubscript{PASS} from 0.9V to 2.5V are applied and then T\textsubscript{2} \textit{V}\textsubscript{TH} is measured. Similar to the single cell demonstration on single-port FeFET shown in Fig.\ref{fig:Fig2}\textbf{k}, severe disturb can happen for a large \textit{V}\textsubscript{PASS}. For example, if \textit{V}\textsubscript{PASS}=2.3V, then 100 $\mu$s pass time can completely flip the memory state, thus posing a serious concern over the state stability. This disturb is of course highly stress bias dependence, as also seen in the read disturb characterization to T\textsubscript{1}\&T\textsubscript{3} shown in Fig.\ref{fig:Fig3}\textbf{e}, where the state is not disturbed.    

\begin{figurehere}
  \centering
    \includegraphics[scale=1,width=0.95\textwidth]{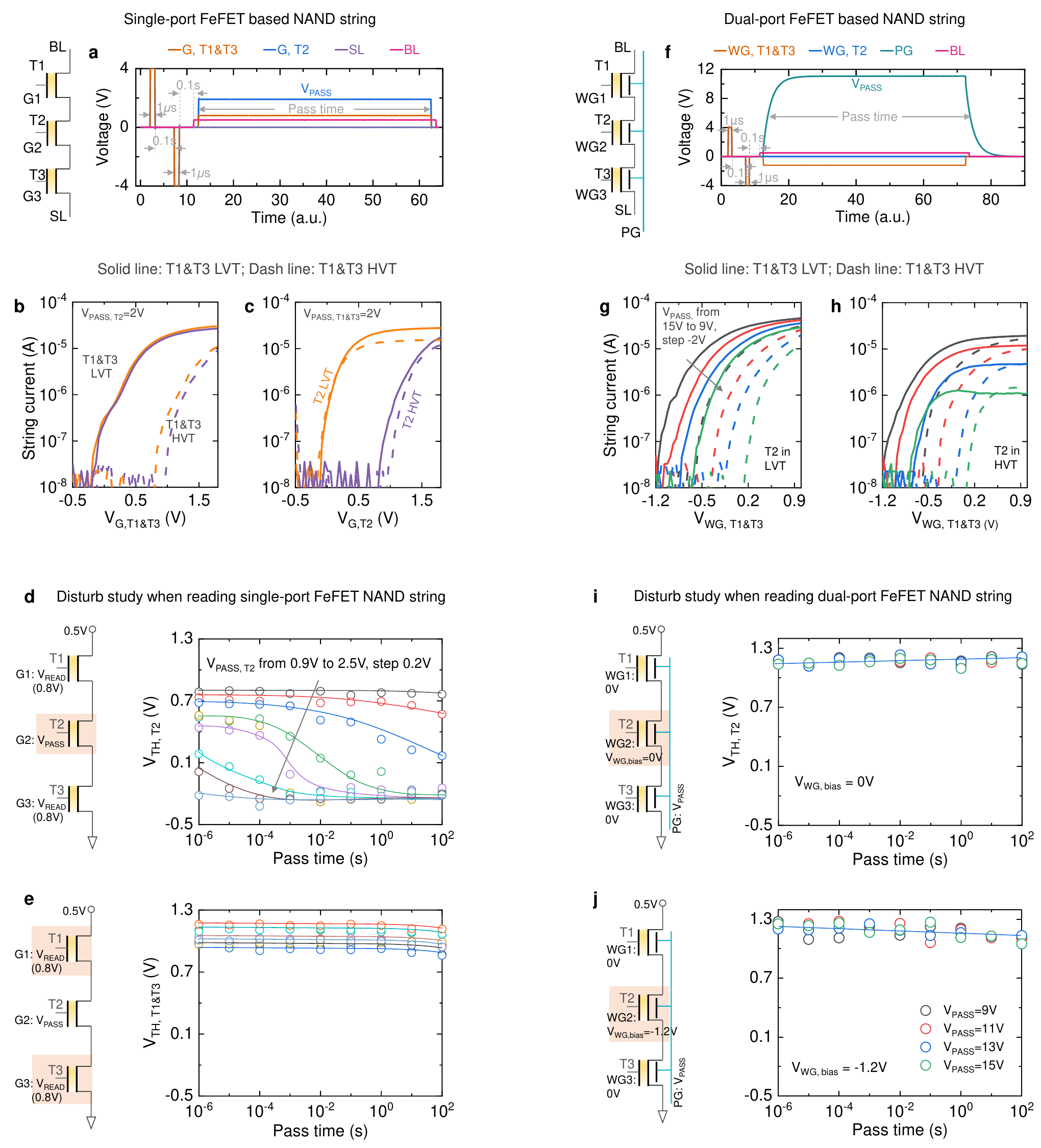}
    \captionsetup{parbox=none} % for caption to be split
    \caption{\textit{\textbf{Experimental validation of the pass disturb-free operation in the NAND FeFET string.} \textbf{a.} Experimental setup and waveforms for the single-port FeFET based NAND string. The gates of T1 and T3 are wired together for the ease of characterization. \textbf{b-c.} String current when respectively sweeping \textit{V}\textsubscript{G, T1\&T3} and \textit{V}\textsubscript{G, T2} with \textit{V}\textsubscript{PASS}=2 V. \textbf{d.} Characterization of the pass disturb under different \textit{V}\textsubscript{PASS} conditions. A higher \textit{V}\textsubscript{PASS} causes a shorter pass time for the \textit{V}\textsubscript{TH} state to be disturbed. \textbf{e.} Read disturb is unnoticeable in T1 and T3 as \textit{V}\textsubscript{READ} is much lower than \textit{V}\textsubscript{PASS}. \textbf{f.} Experimental setup and waveforms for the dual-port FeFET based NAND string. \textbf{g-h.} String current sweep when applying the pass voltage on the PG of T2 for the LVT and HVT state, respectively. A higher \textit{V}\textsubscript{PASS} is necessary to achieve a high string current when T2 is in the HVT state. \textbf{i.} No \textit{V}\textsubscript{TH} shift is found when reading the NAND string from the PG. \textbf{j.} The condition of \textit{V}\textsubscript{WG, bias}=-1.2 V is also measured to confirm the pass disturb-free operation.}}
    \label{fig:Fig3}
\end{figurehere}

Dual-port FeFET can eliminate pass disturb by incorporating separate ports for write/read and pass operations. Fig.\ref{fig:Fig3}\textbf{f} shows the three-transistor NAND string and the corresponding testing waveforms. The p-well body contact shared among the three transistors are used for pass operation. The waveform shows write and read pulses of T\textsubscript{1}\&T\textsubscript{3} and T\textsubscript{2} write gate is grounded while the pass bias is applied on the pass gate. Fig.\ref{fig:Fig3}\textbf{g} and \textbf{h} show the \textit{I}\textsubscript{D}-\textit{V}\textsubscript{G} characteristics of T\textsubscript{1}\&T\textsubscript{3} when the unselected cell T\textsubscript{2} is at LVT state and HVT state, respectively. It shows that sensing of the target cell memory state can be successfully realized. Note that the results also show that a high enough \textit{V}\textsubscript{PASS} is required otherwise the HVT state device is not fully turned ON that could limit the string current, as shown in Fig.\ref{fig:Fig3}\textbf{h}. Fig.\ref{fig:Fig3}\textbf{i} and \textbf{j} study pass disturb in the dual-port NAND FeFET string. The \textit{V}\textsubscript{TH} of cell T\textsubscript{2} is measured after the pass bias is applied on the pass gate. The results show that T\textsubscript{2} is not disturbed at all even when a \textit{V}\textsubscript{PASS}=15V is applied. These results therefore confirm the pass disturb free nature of the dual-port NAND string.
\section*{\large Scaled Dual-Port Vertical NAND FeFET Operation and Integration}
Previous verifications on planar dual-port FeFET device and NAND string have demonstrated that the pass disturb free operation originates from its unique structural property. Such a principle should also be applicable for vertical NAND array. To verify the dual-port operation in a practical vertical NAND string, TCAD simulations are performed. The cross section of a dual-port memory cell and its parameters are shown in Fig.\ref{fig:Fig4}\textbf{a}. A 3D model of a dual-port NAND string with 8 WLs is shown in Fig.\ref{fig:Fig4}\textbf{b}. In the simulation, operations to write and read WL\textsubscript{3} cell is shown in Fig.\ref{fig:Fig4}\textbf{c}. First, all the cells are erased, after which the WL\textsubscript{3} cell is read out. To do that, the center pass gate is applied a pass bias and a WL\textsubscript{3} read bias is applied. In this case, a low string current is read out due to the erase operation. Then WL\textsubscript{3} cell is programmed into the LVT state. After the read operation, a high string current is sensed. Therefore, it demonstrates the feasibility of the proposed technique in vertical NAND FeFET string. In addition, Fig.\ref{fig:Fig4}\textbf{d} and \textbf{e} show the electric field in the gate stack for both single- and dual-port string for the pass cell in the HVT state. When \textit{V}\textsubscript{PASS} is applied also on the write gate, the depolarization field in the ferroelectric is enhanced while it is reduced when the pass voltage is applied on the central pass gate in dual-port FeFET string. The extracted electric field distribution of a NAND string with different \textit{V}\textsubscript{PASS} is shown in the  Supplementary Fig.\ref{fig:sups2}. These results demonstrates that pass disturb free operation is applicable for vertical NAND array. 

% This is consistent with the electric field distribution in the NAND string (Fig.6(d)).  It is also worth noting that the proposed design is compatible with existing vertical NAND with negligible overhead. A bottom pass gate contact patterned in the bottom substrate is proposed to connect all the pass gates in a block (Fig.6(e)). The top view and cutline views in Fig.6 (f)-(h) show more details about gate interconnections. Leveraging the global pass gate, the pass gate interconnection is possible with negligible overhead and the pass operation can be performed in the disturb-free way.

It is also worth noting that the proposed design is compatible with existing vertical NAND process integration with minimum overhead \cite{aritome2015nand, prince2014vertical}. Supplementary Fig.\ref{fig:sups3}{} shows a tentative process integration flow to integrate a center core pass gate. The center core metal can be filled after memory hole etching and subsequent deposition of ferroelectric, poly-silicon channel, and pass gate non-ferroelectric dielectric. A few other steps, such as selectively deposition of the gate metal for the string select transistor, can be included for complete processing. The resulting 3D structure is shown in Fig.\ref{fig:Fig4}\textbf{f}. It suggests a similar structure to conventional vertical NAND array \cite{park2014three}. However, for each string, the central filler is no longer a dielectric, but a metallic pass gate, which is connected to a bottom pass gate contact patterned in the bottom substrate in a block. The top view and cutline views in Fig.\ref{fig:Fig4}\textbf{g}-\textbf{i} show more details. Leveraging the global pass gate, the pass operation can be performed with a low cost.

\begin{figurehere}
  \centering
    \includegraphics[scale=1,width=0.95\textwidth]{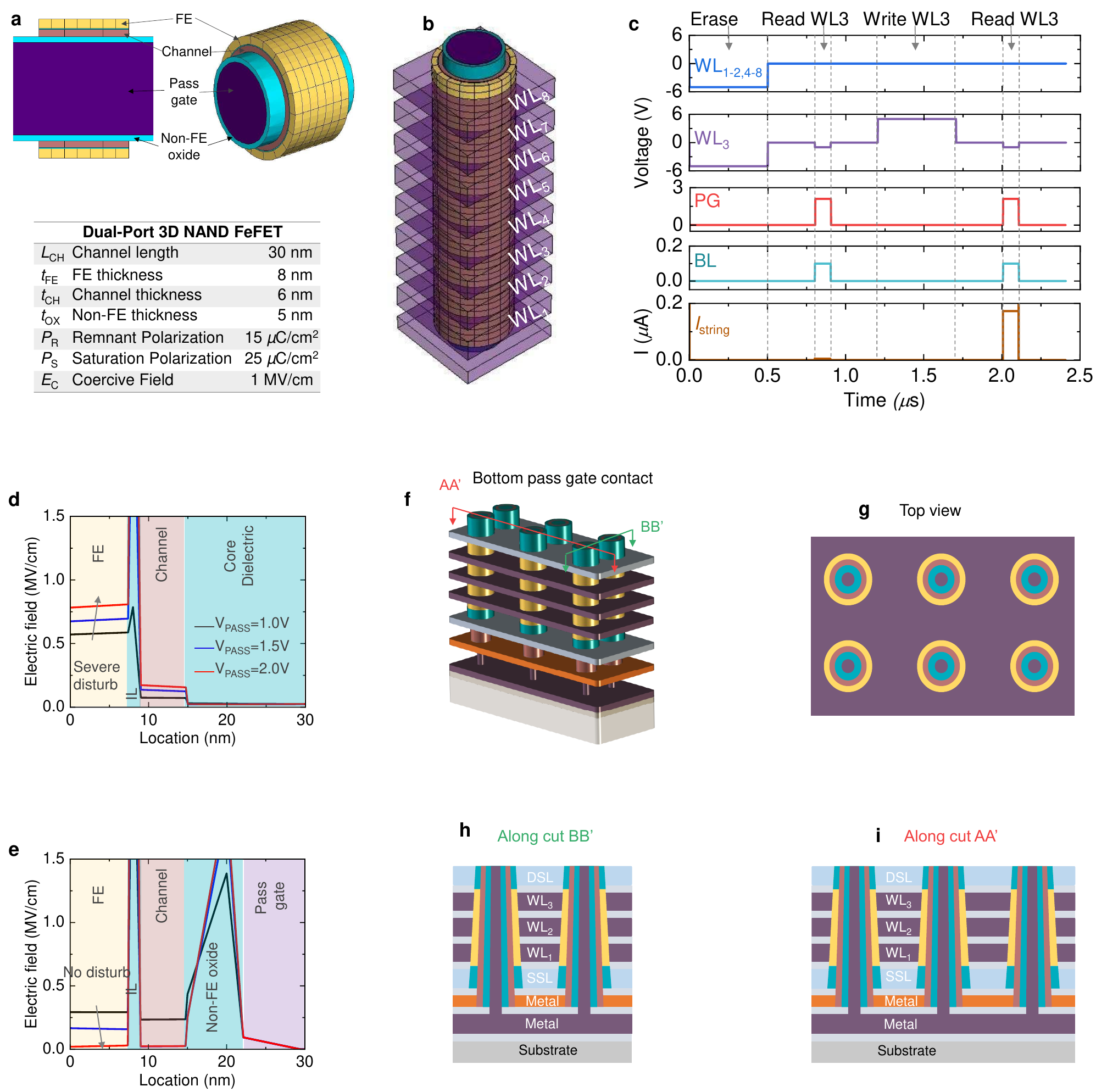}
    \captionsetup{parbox=none} % for caption to be split
    \caption{\textit{\textbf{Scaled dual-port vertical NAND FeFET operation and integration.} \textbf{a.} Cross-section of the dual-port memory cell and its parameters in TCAD simulation. \textbf{b.} 3D model of the dual-port NAND string with 8 WLs. \textbf{c.} Operation waveforms showing erase, write, and read operations. WL\textsubscript{3} is the select cell. \textbf{d-e.} Electric field distribution in the single- and dual-port gate stack of the pass cell in the HVT state. The electric field induced by \textit{V}\textsubscript{PASS} enhances the depolarization field in the single-port string while degrades it in the dual-port string. \textbf{f.} Proposed 3D design for the vertical NAND compatible structure. It features a core pass gate in the center of the string with a pass gate contact in the bottom. \textbf{g-i.} Top view and cutline views of the proposed design showing the details of the global pass gate.}}
    \label{fig:Fig4}
\end{figurehere}
\section*{\large Conclusion}
We have demonstrated the challenges of the single-port NAND FeFET in handling pass disturb and the effectiveness of the proposed dual-port design. We also explored our design in highly scaled vertical NAND memory through TCAD and proposed a global bottom pass gate contact. These demonstrations indicate that the proposed design is very promising in enabling high-reliability dense storage.

\section*{\large Data availability}
The data that support the plots within this paper and other findings of this study are available from the corresponding author on reasonable request.

\section*{\large References}
\bibliography{ref.bib}

\section*{\large Acknowledgements}

This work was partly supported by SUPREME and PRISM, two of the JUMP 2.0 centers, NSF 2312884, and the United States Air Force under Air Force Contract No. FA8702-15-D-0001. Any opinions, findings, conclusions, or recommendations expressed in this material are those of the author(s) and do not necessarily reflect the views of the United States Air Force. This work is funded by the German Bundesministerium für Wirtschaft (BMWI) and by the State of Saxony in the frame of the “Important Project of Common European Interest (IPCEI)”.

\section*{\large Author contributions}

K.N., V.N., and S.D. proposed and supervised the project. Z.Z., Z.J., S.D., and Y.X. performed FEOL FeFET experimental verification. K.A.A. performed BEOL FeFET experimental verification. S.W. and S.Y. conducted 3D TCAD simulations. H.M., S.D., D.K., S.S., and S.B. fabricated the FEOL FeFET devices. R.J. helped with the device validation. S.M., and M.M. helped with the FDSOI FeFET operations as a dual-port FeFET. S.L., K.K., K.K., W.K., and D.H. helped discussion of the project. All authors contributed to write up of the manuscript. 

\section*{\large Competing interests}
An invention disclosure has been submitted through the Pennsylvania State University.

\newpage

\renewcommand{\thefigure}{S\arabic{figure}}
\renewcommand{\thetable}{S\arabic{table}}
\setcounter{figure}{0}
\setcounter{table}{0}

\newpage
\begin{center}
\title{\textbf{\Large Supplementary Materials}}
\end{center}

\vspace{-4ex}
\section*{\large Device Fabrication}
In this paper, the fabricated fully depleted silicon-on-insulator (FDSOI) ferroelectric field effect transistor (FeFET) features a poly-crystalline Si/TiN/doped HfO\textsubscript{2}/SiO\textsubscript{2}/p-Si gate stack. The devices were fabricated using a 22 nm node gate-first high-$\kappa$ metal gate CMOS process on 300 mm silicon wafers. The buried oxide is 20nm SiO\textsubscript{2}. Detailed information can be found in \cite{dunkel2017fefet, mulaosmanovic2021ferroelectric}. The ferroelectric gate stack process module starts with growth of a thin SiO\textsubscript{2} based interfacial layer, followed by the deposition of the doped HfO\textsubscript{2} film via atomic layer deposition (ALD). 
A TiN metal gate electrode was deposited using physical vapor deposition (PVD), on top of which the poly-Si gate electrode is deposited. The source and drain n+ regions were then activated by a rapid thermal annealing (RTA) at approximately 1000 $^\circ$C. This step also results in the formation of the ferroelectric orthorhombic phase within the doped HfO\textsubscript{2}.

\newpage
\section*{\large Electrical Characterization}

The experimental verification was performed with a Keithley 4200-SCS Semiconductor Characterization System (Keithley system), a Tektronix TDS 2012B Two Channel Digital Storage Oscilloscope (oscilloscope). Two 4225-PMUs (pulse measurement units) were utilized to generate proper waveforms. In the experimental characteristics, all signals were generated by the Keithley system. The drain/string currents were also captured by the Keithley system. The \textit{V}\textsubscript{TH} was extracted with a constant current of $I_D=10^{-7} W/L $ \textit{A}.

\newpage
\section*{\large Characterization of the Dual-Port BEOL IWO FeFET}
\begin{figurehere}
  \centering
    \includegraphics[width=1\textwidth]{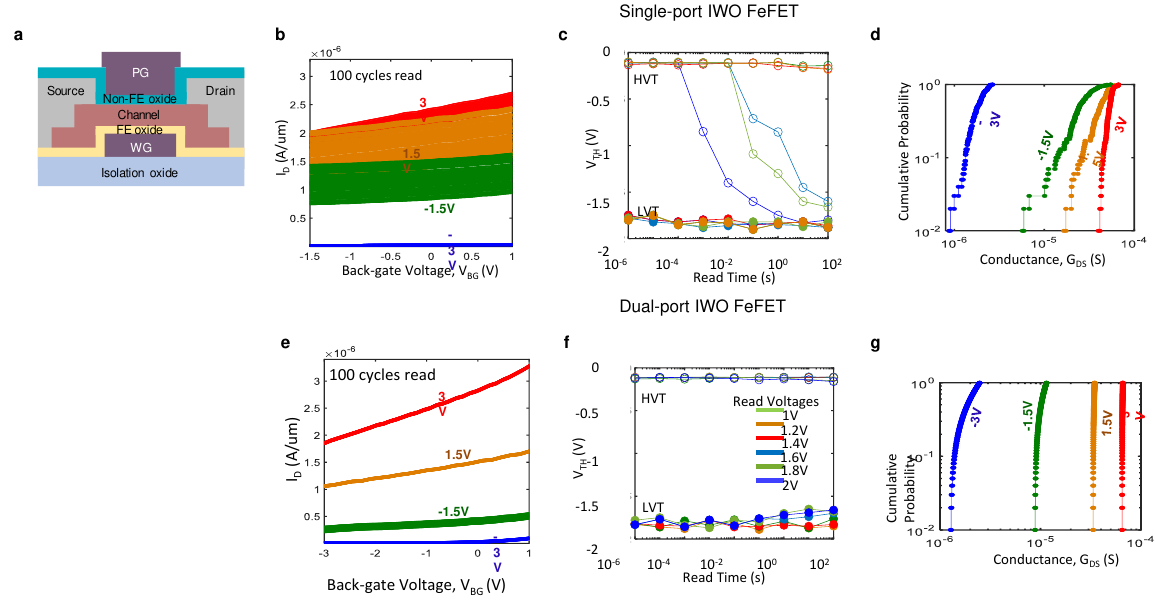}
    \captionsetup{parbox=none} % for caption to be split
    \caption{ \textbf{Characterization of the dual-port BEOL IWO FeFET} \textbf{a.} Schematic of the dual-port BEOL IWO FeFET. FE oxide and non-FE oxide are Hf\textsubscript{0.5}Zr\textsubscript{0.5}O\textsubscript{2} and HfO\textsubscript{2}, respectively. \textbf{b\&e.} \textit{I}\textsubscript{D}-\textit{V}\textsubscript{G} curves of the single- and dual-port IWO FeFETs. \textbf{c\&f} \textit{V}\textsubscript{TH} shifts measured after different pass time under different pass voltages for the single- and dual-port IWO FeFETs. \textbf{d\&g} Cumulative cycle-to-cycle drain-source conductance shift of the single- and dual-port IWO FeFETs. }
    \label{fig:sups1}
\end{figurehere}

The pass-disturb free operation is experimentally verified on the dual-port BEOL FeFET. The FeFET is realized with the ferroelectric placed below the amorphous metal oxide thin film channel (e.g., tungsten doped indium oxide (IWO) in this work) and the non-ferroelectric layer placed above the channel. Fig.\ref{fig:sups1}\textbf{a} shows the schematic of the dual-port BEOL IWO FeFET. The ferroelectric layer is Hf\textsubscript{0.5}Zr\textsubscript{0.5}O\textsubscript{2} (HZO) deposited by ALD and the non-ferroelectric layer is HfO\textsubscript{2}. Multi-level cell (MLC) operations with 100-cycle read of the single- and dual-port IWO FeFETs are shown in Fig.\ref{fig:sups1}\textbf{b\&e}. The \textit{V}\textsubscript{TH} shifts measured with different pass time lengths under different pass voltages are shown in Fig.\ref{fig:sups1}\textbf{c\&f}. The pass disturb is clearly observed in the single-port IWO FeFET while the dual-port FeFET exhibits excellent pass disturb-free feature. Fig.\ref{fig:sups1}\textbf{d\&g} show the cumulative cycle-to-cycle drain-source conductance (\textit{G}\textsubscript{DS}) shift of the IWO FeFETs.

\newpage
\section*{\large Extracted Electric Field of the NAND String}
\begin{figurehere}
  \centering
    \includegraphics[width=1\textwidth]{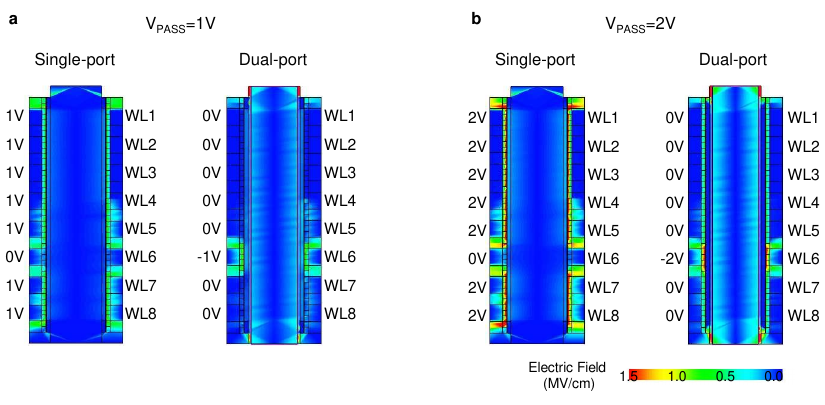}
    \captionsetup{parbox=none} % for caption to be split
    \caption{ \textbf{Extracted electric field of the NAND string from TCAD simulation.} Electric field distribution of the single- and dual-port string for \textbf{a.} \textit{V}\textsubscript{PASS}=1 V and \textbf{b.} \textit{V}\textsubscript{PASS}=2 V. }
    \label{fig:sups2}
\end{figurehere}
Fig.\ref{fig:sups2} shows the extracted electric field with (a) \textit{V}\textsubscript{PASS}=1 V and (b) \textit{V}\textsubscript{PASS}=2 V for both single-port and dual-port string. The electric field is obtained when WL6 is erased to the HVT state. It shows a highly concentrated electric field in the ferroelectric for single-port FeFET string, which is against the polarization.

\newpage
\section*{\large Tentative Process Flow for the Dual-Port Vertical FeFET-based NAND Storage}
\begin{figurehere}
  \centering
    \includegraphics[width=1\textwidth]{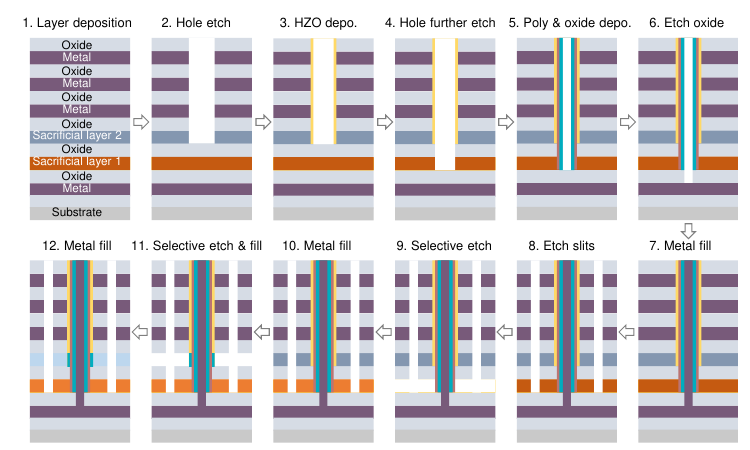}
    \captionsetup{parbox=none} % for caption to be split
    \caption{ \textbf{Tentative process integration for the dual-port vertical FeFET-based NAND Storage.} The center core metal can be filled after memory hole etching and subsequent deposition of ferroelectric, poly-silicon, and PG dielectric.}
    \label{fig:sups3}
\end{figurehere}
Fig.\ref{fig:sups3} shows a tentative process integration flow for the dual-port NAND FeFET with the string-compatible pass gate placed in the center core. The fabrication starts with the layer deposition, hole etch, FE HZO deposition, and then channel and oxide deposition. One important step is the contact metal to the silicon channel (step 7-10) because it is no longer connected with the substrate. The process is plausible and 3D NAND compatible, thus making it a practical design.

\textsubscript{}

\end{document}